\documentclass[conference]{IEEEtran}
\IEEEoverridecommandlockouts
\usepackage{amsmath,amssymb,booktabs,graphicx,xcolor}
\usepackage{microtype}
\usepackage{placeins}
\usepackage[hidelinks]{hyperref}

\newcommand{\Rset}{\mathcal{A}}
\title{Early UAV Detection Using 5G NR LEO Downlinks for Passive ISAC\thanks{Implementation code: \url{https://github.com/mombash/5g-leo-isac}.}}
\author{
\IEEEauthorblockN{Mohamed Hamad\IEEEauthorrefmark{1}, Nizar Zorba\IEEEauthorrefmark{1},
Tamer Khattab\IEEEauthorrefmark{1}, Hamid Saeedi\IEEEauthorrefmark{2}, Halim Yanikomeroglu\IEEEauthorrefmark{3}}
\IEEEauthorblockA{\IEEEauthorrefmark{1}Department of Electrical Engineering, Qatar University, Doha, Qatar}
\IEEEauthorblockA{\IEEEauthorrefmark{2}College of Engineering and Technology, University of Doha for Science and Technology, Doha, Qatar}
\IEEEauthorblockA{\IEEEauthorrefmark{3}Department of Systems and Computer Engineering, Carleton University, Ottawa, Canada}
\IEEEauthorblockA{Emails: \{m.bashir.hamad, nizarz, tkhattab\}@qu.edu.qa, hamid.saeedi@udst.edu.qa, halim@sce.carleton.ca}}
\newcommand{\NeuralAlertTime}{0.997}
\newcommand{\ExhaustiveAlertTime}{1.571}
\newcommand{\DualAlertTime}{1.277}
\newcommand{\NeuralReductionExhaustive}{36.5}
\newcommand{\NeuralReductionDual}{21.9}
\newcommand{\TimingUAVAttempts}{64}
\newcommand{\TimingEmptyAttempts}{32}
\newcommand{\TimingTargetRecords}{16}
\newcommand{\TimingEmptyRecords}{8}

\begin{document}
\raggedbottom
% IEEE preprint notice in the blank first-page header area.
\AddToHookNext{shipout/foreground}{%
  \put(0,-25){\makebox[\paperwidth][c]{%
    \parbox[t]{\textwidth}{\centering\normalfont\small
      This work has been submitted to the IEEE for possible publication. Copyright may be transferred without notice, after which this version may no longer be accessible.\par}}}%
}
\maketitle
\begin{abstract}
5G New Radio (NR) low-Earth-orbit (LEO) downlinks enable passive integrated sensing and communication (ISAC) for unmanned aerial vehicle (UAV) detection. Weak echoes require motion-compensated accumulation, but exhaustive searches postpone alerts. We propose a receiver that combines selective Doppler-rate search, early UAV alerts and continued delay--Doppler estimation. Classical, statistical and neural configurations select rates of Doppler change to search, while a shared correlation score determines detection. Simulated echo delay, Doppler and power follow the moving satellite--UAV--receiver geometry, propagation loss, assumed scattering and antenna response. In a line-of-sight scene with acquired beams and UAVs at 5--8 km horizontal distance, all three configurations match exhaustive search at 92.9\% correct final detections, with 0.10--0.13\% observed false alerts on no-UAV recordings. The neural configuration reduces mean time to the first correct alert by \NeuralReductionExhaustive\% versus exhaustive search and \NeuralReductionDual\% versus the closest literature baseline, averaging over successfully detected UAVs. These results demonstrate earlier UAV alerts while preserving exhaustive search's final detection accuracy, supporting passive sensing over cellular satellites.
\end{abstract}
\begin{IEEEkeywords}
Passive ISAC, passive radar, 5G NR, LEO satellites, UAV detection, motion compensation.
\end{IEEEkeywords}

\section{Introduction and Related Work}
Direct-to-device satellite links extend terrestrial cellular coverage. Starlink's Direct to Cell connects 4G Long Term Evolution (LTE) phones through low-Earth-orbit (LEO) satellites with T-Mobile and One NZ~\cite{spacex_dtc}. AST SpaceMobile has demonstrated LTE/5G links with AT\&T and Vodafone~\cite{ast2025}. 3GPP specifies 5G New Radio (NR) satellite access within non-terrestrial networks (NTNs)~\cite{nr211,nrntn}.

These downlinks provide a standardized waveform for UAV detection through passive integrated sensing and communication (ISAC) without dedicated radar transmission~\cite{leo_passive}. Terrestrial 5G has supported passive vehicle detection~\cite{samczynski}, alongside ISAC research on road-vehicle positioning~\cite{hdv_voi} and UAV sensing-platform optimization~\cite{haps_uav}. Satellite illumination extends coverage, but distant UAV echoes are weak and change delay and frequency during accumulation~\cite{leo_passive}.

A passive receiver compares a direct-signal \emph{reference} recording with a \emph{surveillance} recording containing reflections. Extra travel time produces delay, and motion produces differential Doppler. The receiver aligns delay and phase to accumulate echoes coherently. Longer observations collect more energy but require compensation for Doppler's rate of change. Exhaustive search repeats accumulation across candidate delays, Dopplers and rates~\cite{li}, postponing alerts for echoes already detectable.

\begin{figure}[!b]
\centering\includegraphics[width=\columnwidth,trim=0 3bp 0 4bp,clip]{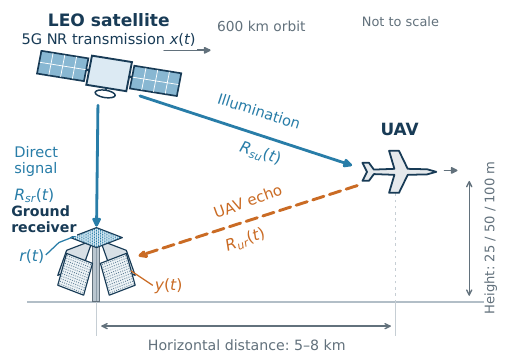}
\caption{Satellite illumination and UAV reflection. Gray arrows show motion.}
\label{fig:scenario}
\end{figure}

Classical search methods have been developed to reduce this computation. In dual-scale detection, changing delay is first corrected on a coarse grid, and that alignment is reused in finer phase-correction searches~\cite{li}. For Global Navigation Satellite System (GNSS) passive radar, keystone alignment and Lv's distribution estimate Doppler and its rate~\cite{he}. Alternatively, two-step sequential detection reserves a second observation for unresolved locations~\cite{grossi}.

Learning has been studied for temporal prediction~\cite{mcst_mamba} and reducing computation in communication and sensing. Learned super-resolution has been used to reduce ray-tracing cost in propagation modeling~\cite{propagation_dt}. For target detection, neural trajectory predictions have been used to restrict generalized Radon--Fourier transform (GRFT)-like motion searches~\cite{wang2024}. In learned feature selection, further measurements are chosen from observed information~\cite{covert}. 

Our contributions are:
\begin{itemize}
\item To our knowledge, we introduce the first simulator for distant UAV detection using 5G NR LEO downlinks that jointly derives echo delay, Doppler and received power from the evolving satellite--UAV--receiver geometry, propagation loss, assumed target scattering and antenna response. Within this framework, we develop a passive receiver combining selective Doppler-rate search, early UAV alerts and continued delay--Doppler estimation.
\item We develop adaptive classical, statistical and neural configurations for rate selection. The learned configurations use initial search scores and phase changes to choose further rates, with shared correlation-based detection and estimation.
\item We compare correct detections, alert time and false alerts with exhaustive search and three literature baselines, and quantify first-to-final estimation-error reduction.
\end{itemize}

Our code includes the simulator, receiver and baseline implementations, feature definitions, model architectures and execution instructions.

\section{System Model and Detection Problem}
\subsection{Physical Setup and Received Signals}
Figure~\ref{fig:scenario} shows the bistatic satellite--UAV--receiver arrangement. An upward array receives the direct reference $r(t)$, while one of four side arrays receives surveillance signal $y(t)$ from the UAV direction. Each array has $12\times12$ elements with half-wavelength spacing, and the side arrays tilt $20^\circ$ upward. For satellite--receiver, satellite--UAV and UAV--receiver distances $R_{sr}(t)$, $R_{su}(t)$ and $R_{ur}(t)$, excess path length $L(t)$ and echo delay $\tau(t)$ are
\begin{align}
 L(t)&=R_{su}(t)+R_{ur}(t)-R_{sr}(t),\label{eq:path}\\
 \tau(t)&=L(t)/c,\label{eq:delay}
\end{align}
where $c$ is the speed of light. Echo generation accounts for motion during transmission, reflection and reception. The detector receives recordings, not these distances, and must estimate delay.

We use the direct satellite path as the common timing and phase reference. After aligning the direct-path components, the complex baseband signals are modeled as
\begin{align}
 r(t)&=b(t)x(t)+n_r(t),\\
 y(t)&=d(t)x(t)+a(t)x(t-\tau(t))e^{-j2\pi f_c\tau(t)}+n_y(t),
 \label{eq:signal}
\end{align}
where $x(t)$ is the transmitted NR waveform, $f_c$ is carrier frequency, and $n_r(t)$ and $n_y(t)$ are independent complex Gaussian noises. Coefficients $b(t)$ and $d(t)$ include propagation loss, antenna gain and phase, and polarization coupling. The term $d(t)x(t)$ is direct-signal leakage. In the echo, $x(t-\tau(t))$ represents extra travel time, $a(t)$ includes two-leg loss, antenna response, polarization and scattering, and $e^{-j2\pi f_c\tau(t)}$ supplies the excess carrier phase that produces differential Doppler. Detection uses noisy reference $r(t)$.

UAV echo power $P_e(t)$ follows the bistatic radar equation~\cite{samczynski},
\begin{equation}
 P_e(t)=\frac{\mathcal{E}(t)G_r(t)\lambda^2\sigma}
 {(4\pi)^3R_{su}^2(t)R_{ur}^2(t)},
 \label{eq:power}
\end{equation}
where $\mathcal{E}(t)$ is satellite equivalent isotropically radiated power (EIRP) toward the UAV, combining transmit power and antenna gain over the 20-MHz channel. Effective receiver gain $G_r(t)$ includes pointing and polarization coupling, and wavelength is $\lambda=c/f_c$. Bistatic radar cross section (RCS) $\sigma$ (m$^2$) is the effective scattering area toward the receiver. It differs from UAV physical area and depends on shape, material, viewing geometry and polarization.

\begin{table}[!t]
\caption{Common waveform, hardware and scene parameters}
\label{tab:system}\centering\footnotesize
\begin{tabular}{@{}p{.47\columnwidth}p{.48\columnwidth}@{}}
\toprule
Parameter & Value or modeling choice\\\midrule
Carrier $f_c$; wavelength $\lambda$ & 2.19 GHz; 0.137 m\\
Channel / occupied bandwidth $B$ & 20 / 18.36 MHz\\
Sample rate $F_s$; subcarrier spacing & 30.72 MHz; 30 kHz\\
Resource blocks; data modulation & 51; fully loaded QPSK\\
Satellite altitude; EIRP density & 600 km; 34 dBW/MHz\\
Aperture efficiency; noise figure & 0.6; 3 dB\\
Antenna temperature; steering refresh & 290 K; 0.1 s\\
UAV horizontal distance; height & 5--8 km; 25, 50, 100 m\\
UAV speed; radar cross section $\sigma$ & 45--80 m/s; $0.10$ m$^2$\\
Propagation & Moving line-of-sight paths\\
Search delay; differential Doppler & 2.60--58.30~$\mu$s; $\pm1600$ Hz\\
Doppler-rate grid & $-52:4:52$ Hz/s (27 candidates)\\
\bottomrule
\end{tabular}
\end{table}

We generate an n256-band 5G signal with quadrature phase-shift keying (QPSK) data using Table~\ref{tab:system} and the specifications~\cite{nr211,nrntn}. The illumination power follows the S-band, 600-km LEO calibration setting in Table~6.1.1.1-1 of 3GPP TR~38.821~\cite{nr821}. Array design and RCS are author-selected. The UAV is modeled as a point scatterer retaining the incident field component along an assumed target polarization axis. Scattering phase is fixed within each flight and random between flights, while propagation phase varies with path length. We assume acquired beams and update their pointing from the modeled satellite and UAV positions every 0.1~s. Common timing and known satellite ephemeris permit reference alignment. 

\begin{figure*}[!t]
\centering\includegraphics[width=\textwidth,trim=0 3bp 0 0bp,clip]{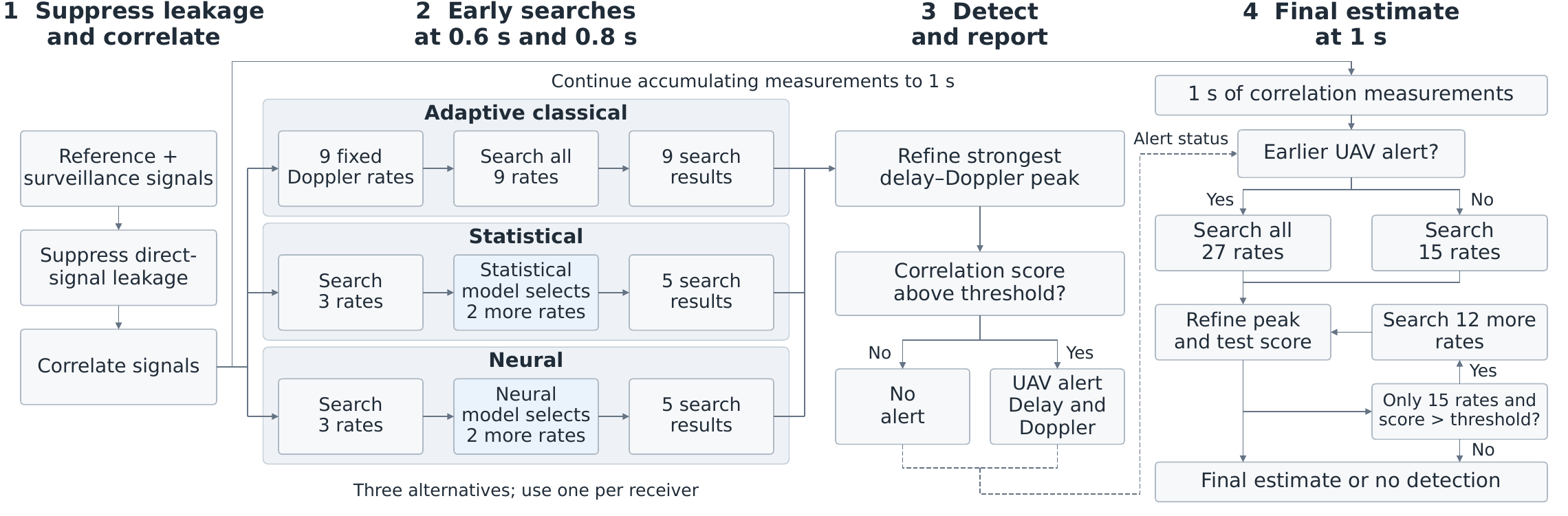}
\caption{Receiver architecture. Each rate search produces a delay--Doppler map. The 0.6-, 0.8- and 1-s labels denote observation durations, before processing finishes.}
\label{fig:architecture}
\end{figure*}

\subsection{Delay, Doppler and Motion Compensation}
The phase in~\eqref{eq:signal} gives differential Doppler and its rate
\begin{equation}
 f_D(t)=-f_c\frac{d\tau(t)}{dt},\qquad
 \alpha(t)=\frac{df_D(t)}{dt}=-f_c\frac{d^2\tau(t)}{dt^2}.
 \label{eq:doppler}
\end{equation}
Doppler $f_D$ has units Hz and its rate $\alpha$ has units Hz/s. The physical delay is unknown to the receiver. Over duration $T$, a constant candidate rate $\alpha$ gives linear Doppler and quadratic delay. Candidate delay $\tau_{\mathrm{mid}}$ and Doppler $f_{\mathrm{mid}}$ are defined at the midpoint. Motion compensation aligns correlation measurements in delay and phase before accumulation. With $t$ measured from observation start, $t-T/2$ is the time relative to that midpoint, and predicted delay is

\begin{equation}
 \widetilde{\tau}(t)=\tau_{\mathrm{mid}}
 -\frac{f_{\mathrm{mid}}}{f_c}(t-T/2)
 -\frac{\alpha}{2f_c}(t-T/2)^2.
 \label{eq:trajectory}
\end{equation}
For $0\leq t\leq T$, the corresponding phase relative to the midpoint is
\begin{equation}
 \widetilde\phi(t)=2\pi\left[f_{\mathrm{mid}}(t-T/2)+\frac{\alpha}{2}(t-T/2)^2\right].
 \label{eq:phase}
\end{equation}

An incorrect rate leaves quadratic phase error that reduces coherent accumulation~\cite{li,he}. Each rate search also tests candidate delays and Dopplers to select and phase-align correlation measurements, retaining their largest score.

\subsection{From Samples to a Detection}
The receiver subtracts estimated surveillance leakage and correlates the residual with delayed reference samples. Correlation measurement $z(\tau,t_n)$ gives their agreement and phase at trial delay $\tau$. Here $t_n$ is the time of the $n$th discrete correlation measurement within duration $T$, whereas $t$ describes continuous motion. Sampling at the predicted delays, removing the predicted phase and accumulating gives the normalized correlation score

\begin{equation}
 S_T(\tau_{\mathrm{mid}},f_{\mathrm{mid}},\alpha)=
 \frac{\left|\sum_{n\in\mathcal{I}_T}z(\widetilde{\tau}(t_n),t_n)
 e^{-j\widetilde{\phi}(t_n)}\right|^2}{V_T},
 \label{eq:score}
\end{equation}
The index set $\mathcal I_T$ contains measurements within duration $T$. Division by summed noise variance $V_T$ makes scores comparable across noise levels. Simulation amplitudes have unit surveillance-noise variance, preserving physical signal-to-noise ratio. Correlation-noise variance follows reference power, estimated in 5-ms intervals at each delay and summed over accumulated measurements.

Exhaustive search maximizes this score over the delay, Doppler and rate grid, then refines the peak between grid points as described in Section~III-A. A refined score above detection threshold $\gamma$ triggers a UAV alert containing the estimated delay and Doppler. Otherwise it reports no detection. Section~IV-A describes threshold calibration using no-UAV recordings.

\section{Proposed Receiver}
\subsection{Leakage Suppression, Correlation and Estimation}
Figure~\ref{fig:architecture} shows our receiver. In each 20-ms block, least squares scales and phase-adjusts the reference to minimize surveillance energy. Its complex multiplier can change linearly within the block. Subtracting that copy suppresses direct-signal leakage. Fast Fourier transform (FFT) correlation of the residual with the reference produces $z(\tau,t_n)$.

Delay interpolation supplies intermediate samples, keystone resampling compensates changing delay, and an FFT evaluates many Dopplers together~\cite{li,he}. One rate search covers the allowed delay--Doppler region for one $\alpha\in\Rset=\{-52,-48,\ldots,52\}$ Hz/s. We append zeros to double the FFT input length, producing a finer Doppler search grid with spacing $1/(2T)$ Hz. Delay spacing is $1/F_s$.

Six refinement iterations test small changes in delay, Doppler and rate, also fitting parabolas to neighboring scores to propose intermediate positions. We retain only changes that increase the measured correlation score $S_T$, within two delay samples, $1/T$ Hz and 4 Hz/s of the initial peak. At the end of the observation, reported delay is $\widehat\tau=\widetilde\tau(T)$ and Doppler is $\widehat f_D=f_{\mathrm{mid}}+\alpha T/2$, adding the Doppler change since the midpoint.

\subsection{Classical, Statistical and Neural Search Selection}
At $T=0.6$ and $0.8$~s, adaptive classical searches nine fixed rates, $-16,-12,\ldots,16$ Hz/s, selected using training data, and refines the strongest peak. Statistical and neural configurations search $-8,-4,0$ Hz/s, then select and search two further rates from $\Rset$. They reuse the initial results and refine the strongest of all five searches, each using all samples collected for that check.

Both models receive 82 features describing correlation-score differences, delay--Doppler peak variation, temporal phase and refinement changes. The neural input adds an 11-value description of each initial search's strongest peak and three temporal correlation sequences at the strongest initial delay estimate and offsets of $\pm1$ sample. The sequences are accumulated into 64 time bins and rotated by the central sequence's accumulated phase, preserving phase differences over time and neighboring delays. Inputs use only the current observation and initial searches.

Our statistical selector ranks rates using linear weights, biases and softmax. Our neural selector projects the 82 features, encodes peak descriptions with attention, and processes correlation sequences with convolution layers. Mean and maximum pooling summarize the encoded peaks and sequences. These outputs join the projected features before a gated recurrent unit and output layer rank rates. Recurrent state resets at each check, retaining no earlier-check information.

We train both selectors to select two additional rates after three initial searches, supervised by all-rate correlation scores. Options $\mathcal J$ comprise unsearched rates and one option representing the strongest completed search. Each option's score $s_j$ defines its target probability $q_j$:
\begin{equation}
 q_j=\frac{e^{s_j/2}}{\sum_{k\in\mathcal J}e^{s_k/2}},\qquad
 \mathcal L(\theta)=-\sum_{j\in\mathcal J}q_j\log p_\theta(j\mid\mathbf u_T).
 \label{eq:training}
\end{equation}
Here $\mathbf u_T$ contains initial-search inputs, and model parameters $\theta$ define probabilities $p_\theta$ over search options. These rank rates, while the refined correlation score $S_T$ determines detection. Minimizing cross entropy $\mathcal L$ favors high-scoring options and similar probabilities for similar scores. Statistical fitting penalizes large weights. At inference, we discard the additional option and select the two highest-ranked unsearched rates.

\subsection{Early Alerts and Conditional Final Search}
Our adaptive classical, statistical and neural configurations check at 0.6, 0.8 and 1~s. A refined correlation score above $\gamma$ triggers a UAV alert with delay and Doppler estimates. Sample collection continues to one second, allowing later checks to update those estimates. The one-second check determines the final output, including withdrawal of an earlier alert.

An early alert requires all 27 rates at one second. With no earlier alert, we search 15 rates ranked highest by a training-derived one-second prior. A strongest refined score $\leq\gamma$ ends the search with no detection. A score $>\gamma$ triggers the remaining 12 searches, then refinement of the strongest of all 27 results.

\section{Experimental Setup and Evaluation}
\subsection{Data and Implementation}
We generate 5G waveforms in MATLAB using 5G Toolbox and implement propagation, receiver processing and model fitting in Python with PyTorch GPU processing. We train statistical and neural models on 35,712 examples from 216 independent UAV flights and select models and search settings using 4,608 examples from 32 separate development flights. Each training example pairs features from initial searches over 0.6--0.9~s with probability targets derived from all-rate scores. Each flight's RCS and noise realizations remain in one split. Flights contribute equally to training loss. Training uses UAV-present examples to learn rate selection.

We fit neural and statistical models using AdamW and L-BFGS, respectively. Models, normalization, search settings and thresholds are fixed before testing.

Testing uses 512 independent four-second UAV flights, giving 2,048 nonoverlapping one-second detection attempts. We vary speed and lateral acceleration along approaching, receding, crossing and oblique paths. Initial horizontal distances are 5.2--7.65~km, remaining within 5--8~km, with speeds and heights specified in Table~\ref{tab:system}. Satellite elevations span approximately $36^\circ$--$84^\circ$, with varied azimuths and orbit directions. We check that delay, Doppler and rate stay within search ranges and arrival directions within steering limits, without selecting by detection outcome. All receivers use the same recordings from a fixed library of nineteen 5G waveforms with different cell identities, data and control signals. Results are conditional on this waveform library.

False-alert testing uses 3,000 independent one-second no-UAV recordings with noise, satellite geometry and receiver pointing conditions sampled independently for each recording. Each recording uses a waveform selected from a fixed library of nineteen waveforms. Timing continuously replays \TimingTargetRecords{} UAV and \TimingEmptyRecords{} no-UAV four-second recordings, with four nonoverlapping attempts each.

We calibrate $\gamma$ on 240 one-second observations from 60 independent no-UAV recordings using the adaptive classical 0.6/0.8/1-s schedule. Each contributes its largest refined score across the 9/9/27-rate checks. The 99th percentile, $\gamma=28.1487$, equals the separately calibrated one-second 27-rate threshold. Statistical, neural, exhaustive classical and dual-scale search share this dimensionless threshold, although their achieved false-alert rates can differ.

We time receivers separately on an NVIDIA A10-24Q GPU with 24~GB memory, releasing samples at acquisition rate. Suppression and correlation use one GPU stream alongside two search workers. Later checks can start before earlier checks finish, with worker waiting included. Statistical and neural inference use CUDA graph replay to reduce launch overhead.

Timing starts at the observation's first digital sample and includes sample arrival, CPU-to-GPU transfer, suppression, correlation, input preparation, inference, search and refinement. Radio-frequency reception, analog-to-digital conversion and beam/reference acquisition are excluded. Recordings and models are preloaded.

\begin{table*}[!t]
\caption{Detection quality on 2,048 UAV attempts. False alerts on 3,000 no-UAV recordings show observed percentages and one-sided 95\% upper bounds. Timing uses \TimingUAVAttempts{} UAV and \TimingEmptyAttempts{} no-UAV attempts, averaging first-correct UAV reports and weighting class means equally.}
\label{tab:main}\centering\small
\begin{tabular*}{\textwidth}{@{\extracolsep{\fill}}lcccccc@{}}
\toprule Receiver & Final correct & UAV & Combined mean & No-UAV decision & \multicolumn{2}{c}{No-UAV false alerts (\%)}\\ & detections (\%) & alert time (s) & time (s) & time (s) & Observed & 95\% upper bound\\\midrule
Ours: neural & 92.9 & 0.997 & 1.200 & 1.373 & 0.10 & 0.26 \\
Ours: statistical & 92.9 & 1.013 & 1.208 & 1.375 & 0.13 & 0.30 \\
Ours: adaptive classical & 92.9 & 1.043 & 1.223 & 1.375 & 0.13 & 0.30 \\
\midrule
Exhaustive classical & 92.9 & 1.571 & 1.572 & 1.570 & 0.20 & 0.39 \\
Dual-scale search & 93.1 & 1.277 & 1.274 & 1.271 & 0.27 & 0.48 \\
\midrule
Two-step: fixed look & 91.0 & 1.567 & 1.576 & 1.582 & 1.20 & 1.58 \\
Two-step: early look & 85.1 & 1.287 & 1.292 & 1.294 & 0.77 & 1.08 \\
\midrule
Hybrid integration & 91.4 & 4.520 & 4.476 & 4.471 & 0.50 & 0.77 \\
\bottomrule\end{tabular*}

\end{table*}

\begin{figure*}[!t]
\centering\includegraphics[width=\textwidth,trim=0 0bp 0 1bp,clip]{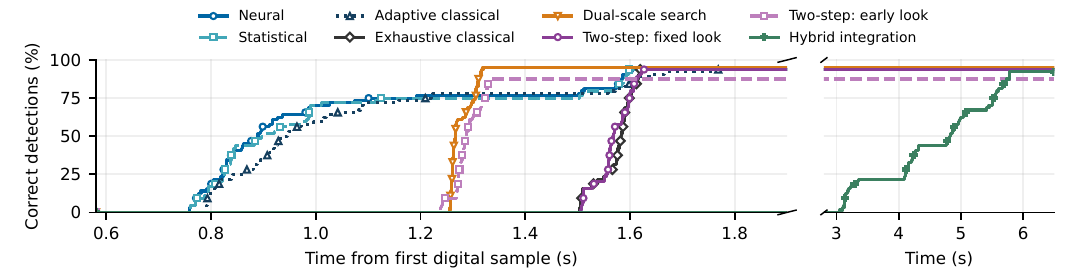}
\caption{Percentage of \TimingUAVAttempts{} UAV timing attempts with a correct delay--Doppler report versus time from the first digital sample. The denominator includes attempts that never produce a correct report.}
\label{fig:time}
\end{figure*}

\subsection{Comparison Methods}
All baselines use the same recordings, hardware and search ranges. \textit{Exhaustive classical} searches all 27 rates after one second and refines the strongest peak. \textit{Dual-scale search} follows~\cite{li}, applying coarse keystone correction once and reusing its output across nearby rates. Residual phase correction precedes accumulation at each rate. It tests all 27 rates with the same refinement as exhaustive search.

We implement the \textit{two-step detection} method in~\cite{grossi}. Upper and lower score thresholds determine detection, rejection or further observation at each delay--Doppler location. Undecided locations are tested again using only new samples. Using development data, we select two settings: a single one-second check (\emph{fixed look}), and a 0.8-s check followed, if needed, by the next 0.2~s (\emph{early look}). Our implementation includes motion compensation between checks and peak refinement for delay--Doppler estimation.

We adapt the \textit{hybrid integration} method in~\cite{he} to one second of 5G samples. Keystone resampling aligns changing delay, then Lv's distribution estimates Doppler and its rate. Peak interpolation refines the estimate. Our scenario uses no GNSS signals, so GNSS-specific processing is omitted, as is integration across multiple one-second observations. We replace the source's constant-false-alarm-rate detector with an independently calibrated maximum-score threshold at a nominal 1\% false-alert probability per observation.

\subsection{Metrics}
A correct detection is an alert with delay error $\leq1/B=54.5$~ns and Doppler error $\leq1/T=1$~Hz, evaluated at the endpoint of that check's samples. The $1/T$ tolerance uses nominal $T=1$~s for both early and final reports.

Table~\ref{tab:main} gives \emph{final correct detections}, the percentage of UAV attempts meeting both limits at the final check, and \emph{UAV alert time}, time to the first correct report averaged over attempts with at least one correct report. Correctness uses simulator ground truth. Figure~\ref{fig:time} shows the percentage of all UAV timing attempts detected correctly by each time. On no-UAV attempts, any alert counts as one false alert, even if withdrawn.

\emph{Combined mean time} weights UAV and no-UAV class means equally, so class proportions do not determine latency performance. UAV attempts contribute first-correct report time, or final decision time if never correct. No-UAV attempts contribute final decision time. Timing uncertainty resamples whole four-second flights, keeping their four attempts together and pairing each flight across receivers.

\section{Results and Discussion}
\subsection{Detection Quality and Response Time}
Table~\ref{tab:main} shows that all three proposed configurations match exhaustive search at 92.9\% correct final detections, versus 93.1\% for dual-scale search, the closest literature baseline. Mean neural UAV alert time is \NeuralAlertTime~s, \NeuralReductionExhaustive\% below exhaustive search's \ExhaustiveAlertTime~s and \NeuralReductionDual\% below dual-scale search's \DualAlertTime~s.

The paired 95\% confidence intervals include zero, so the neural and statistical alert-time reductions relative to adaptive classical are not statistically significant in this measurement session.

Each of our configurations detects correctly in 1,908/2,048 attempts (93.2\%) at some check, but five end without a correct final detection, leaving 1,903 correct final detections (92.9\%). Incorrect first estimates occur in 51 neural and adaptive classical attempts (2.49\%) and 56 statistical attempts (2.73\%). Of these, 50, 50 and 55, respectively, have correct final estimates. Later checks do not guarantee a correct final detection.

The two-step early-look baseline is less accurate than fixed look. Its second check uses only the remaining 0.2~s, whereas fixed look uses the full second. Hybrid integration's products across time lags add processing work, while its queued searches postpone decisions on consecutive observations.

Neural emits 3/3,000 false alerts, versus 4/3,000 for statistical and adaptive classical, each including one withdrawn alert. One-sided 95\% upper bounds are 0.26\% and 0.30\%, respectively.

Neural no-UAV decisions take 1.373~s versus dual-scale search's 1.271~s. Equal class weighting gives a neural combined mean of 1.200~s, 5.8\% below dual-scale search's 1.274~s.

\begin{figure}[!t]
\centering\includegraphics[width=\columnwidth,trim=0 3bp 0 3bp,clip]{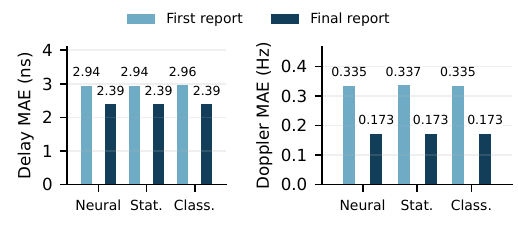}
\caption{Mean absolute delay and Doppler errors of first and final reports on the same 1,903 attempts with correct final detections across all three configurations.}
\label{fig:refinement}
\end{figure}

\subsection{Processing Savings and Later Refinement}
Table~\ref{tab:component} compares full and conditional final search over 16 attempts (four four-second recordings, two per class). Per-check times exclude sample collection and suppression/correlation, which can overlap with search, so their sum is not alert time.

\begin{table}[!t]
\caption{Mean selection, search and refinement time per check (ms).}
\label{tab:component}\centering\footnotesize
\begin{tabular}{@{}lrrr@{}}\toprule Stage & Neural & Statistical & Adaptive classical\\\midrule
0.6-s check & 69.4  & 67.6 & 104.9\\
0.8-s check & 82.7  & 81.5 & 135.6\\
\midrule
1-s check: full search & 474.2  & 477.6 & 465.6\\
1-s check: conditional search & 365.9  & 367.3 & 372.2\\
\bottomrule\end{tabular}

\end{table}

Conditional search saves 22.8\% (neural), 23.1\% (statistical) and 20.0\% (adaptive classical) of one-second-check time.

Figure~\ref{fig:refinement} compares first and final mean absolute errors (MAEs) against truth on the same 512 recordings as Table~\ref{tab:main}, including only correct final detections. Of 2,048 UAV observations, we exclude 143 without final alerts and two with incorrect final alerts, retaining incorrect first estimates. Later checks reduce neural Doppler MAE by 48.5\% (0.335 to 0.173~Hz) and delay MAE by 18.9\% (2.94 to 2.39~ns), using unrounded means.

The configurations have different early searches but the same 27-rate final search and full-second measurements. Refinement starts from the same strongest peak, giving identical final estimates. This analysis excludes 15-rate rejections. Later searches and refinement reduce the mean errors but can increase individual errors. Neural delay error increases in 615 observations (32.3\%) and Doppler error in 486 (25.5\%). Some observations have both increases, but all included final estimates meet both error limits.

\section{Conclusion}
In this work, we developed a passive 5G NR LEO receiver with adaptive classical, statistical and neural configurations for selective Doppler-rate search, early alerts and continued estimation. All three matched exhaustive search at 92.9\% correct final detections for UAVs 5--8~km away. Among correctly detected UAVs, neural mean alert time was \NeuralReductionExhaustive\% shorter than exhaustive search and \NeuralReductionDual\% shorter than the closest literature baseline. Later checks reduced mean errors on correct final detections, supporting timely sensing over cellular satellites without a dedicated radar transmitter. Results assume line-of-sight propagation, direct-signal leakage and acquired beams/reference alignment. Future work includes beam scanning to locate UAVs, acquisition costs, buildings, ground reflections, time-varying scattering and standardized 5G pilots for detection.

\FloatBarrier

\end{document}